\documentclass[twocolumn,aps,prl,amsmath,amssymb]{revtex4}
\usepackage{graphicx}
\usepackage{dcolumn}
\usepackage{bm}
\usepackage[dvips]{color}
\usepackage{epstopdf}
\DeclareGraphicsExtensions{.pdf,.eps,.png,.jpg,.mps}
\begin{document}

\author{Ilya Goykhman$^1$}
\author{Ugo Sassi$^1$}
\author{Boris Desiatov$^2$}
\author{Noa Mazurski$^2$}
\author{Silvia Milana$^1$}
\author{Domenico de Fazio$^1$}
\author{Anna Eiden$^1$}
\author{Jacob Khurgin$^3$}
\author{Joseph Shappir$^2$}
\author{Uriel Levy$^2$}
\author{Andrea C. Ferrari$^1$}
\email{acf26@eng.cam.ac.uk}
\affiliation{$^1$Cambridge Graphene Centre, University of Cambridge, 9 JJ Thomson Avenue, Cambridge CB3 OFA, UK}
\affiliation{$^2$Department of Applied Physics, The Benin School of Engineering and Computer Science, The Hebrew
University, Jerusalem, 91904, Israel}
\affiliation{$^3$Department of Electrical and Computer Engineering, Johns Hopkins University, Baltimore, Maryland
21218, USA}
\title {On-chip integrated, silicon-graphene plasmonic Schottky photodetector, with high responsivity and avalanche photogain}
\begin{abstract}
We report an on-chip integrated metal-graphene-silicon plasmonic Schottky photodetector with 85mA/W responsivity at 1.55$\mu$m and 7\% internal quantum efficiency. This is one order of magnitude higher than metal-silicon Schottky photodetectors operated in the same conditions. At a reverse bias of 3V, we achieve avalanche multiplication, with 0.37A/W responsivity and avalanche photogain$\sim2$. This paves the way to graphene integrated silicon photonics.
\end{abstract}
\maketitle
Over the past decade silicon photonics\cite{ReedSiPhoton2008} has progressed towards miniaturization and on-chip integration of optical communication systems, where data are encoded by light signals and distributed over waveguides, rather than conventional metal-based electronic interconnects\cite{NatPhotonFocus2010, KnightsLPRev2012}. So far, a variety of passive and active photonic devices in Si have been demonstrated, including low-loss ($\sim$0.3dB/cm) waveguides\cite{CardenasOE2009, DesiatovOE2010}, high-quality factor optical cavities ($\sim$10$^6$)\cite{LalanneOE2007,NotomiOE2010,naiman}, high-speed (tens of GHz)\cite{LipsonNat2005, PanicciaOE2007, ReedNP2010} electro-optic modulators and Si light sources based on Raman gain\cite{PanicciaNat2005, BowersNP2010}. The wealth of devices, together with the well established complementary metal-oxide-semiconductor (CMOS) fabrication processes make Si photonics a promising technology for short range (board-to-board, chip-to-chip or intra-chip)\cite{ReedSiPhoton2008} optical communications.

The photodetector (PD) is one of the basic building blocks of an opto-electronic link, where it performs optical-to-electrical signal conversion. Development of Si PDs for telecom wavelengths (1.3-1.6$\mu$m) based on the mature CMOS technology is an essential step for monolithic, on-chip, opto-electronic integration\cite{ReedSiPhoton2008}. While Si PDs are widely employed in the visible spectral range\cite{Sze2006}(0.4-0.7$\mu$m), they are not suitable for detecting near-infrared (NIR) radiation above 1.1$\mu$m, because the energy of NIR photons at telecom wavelengths (0.78-0.95eV) is not sufficient to overcome the Si bandgap (indirect, 1.12eV) and induce photogeneration of electron-hole (e-h) pairs, i.e no photocurrent ($I_{ph}$) is generated. Over the years, the Si photonics industry has developed solutions to overcome this deficiency by combining Ge (bandgap 0.67eV) with Si\cite{CampbellNP2009,KimerlingNP2010, VlasovNat2010} and integrating compound (III-V) semiconductors on the Si chip\cite{HawkinsAPL1997,KangPTL2002} using wafer bonding techniques\cite{BowersRev2010}. While these approaches provide a path towards photodetection in the telecom spectral range\citep{ReedSiPhoton2008}, they either require advanced and complex fabrication processes in the case of SiGe devices\cite{WangRev2011}, or rely on III-V materials systems not compatible with standard CMOS technology\cite{Sze2006}. Motivated by the need of developing Si based PDs for telecom wavelengths several approaches were proposed to date. These include two photon absorption (TPA)\cite{LiangAPL2002, NotomiAPL2010}, defect mediated band-to-band photogeneration via mid-bandgap localized
states\cite{KnightsAPL2005, KnightsJNP2011, DesiatovAPL2014}, deposition of polysilicon\cite{PrestonOL2011} for NIR absorption, enhancement by optical cavities\cite{NotomiAPL2010, KnightsJNP2011, DesiatovAPL2014,PrestonOL2011, PoonAPL2009, CasalinoOL2012}. However, in the cases of defect-mediated and poly-Si PDs, the overall concentration of defects in the Si lattice affects both $I_{ph}$ and the leakage (dark) current $I_{dark}$\cite{KnightsAPL2005, KnightsJNP2011,Sze2006}, i.e a higher defects density increases both the sub-bandgap optical absorption and thermal generation processes\cite{Sze2006}, thus increasing both $I_{ph}$ and $I_{dark}$\cite{Sze2006,KnightsAPL2005, KnightsJNP2011}. As a result, PDs with reduced defects concentration are typically needed\cite{KnightsAPL2005, KnightsJNP2011}, coupled to optical resonators to amplify the optical power and to enhance the absorption without increasing either device length or defect density. On the other hand, nonlinear optical process, such as TPA, could potentially contribute to all-Si NIR-PDs\cite{ReedSiPhoton2008}, but this approach requires increased optical power\cite{NotomiAPL2010} with respect to linear absorption, or PD integration with high quality factor cavities to achieve enhanced photon density\cite{NotomiAPL2010}.

An alternative exploits internal photoemission (IPE) in a Schottky diode\cite{Sze2006, KwongPTL2008, CasalinoAPL2010}. In this configuration, photoexcited ("hot") carriers from the metal are emitted to Si over a potential $\Phi_B$, called Schottky barrier (SB), that exists at the metal-Si interface\cite{Peters1967,
Sze2006}. In Si, the injected carriers are accelerated by an electric field in the depletion region of a Schottky diode and then collected as a photocurrent at the external electrical contacts. Typically, a SB is lower (0.2-0.8eV) than the Si bandgap\cite{Sze2006}, thus allowing photodetection of NIR photons with energy $h\nu>\Phi_B$. The advantages of Schottky PDs are the simple material structure, easy and inexpensive fabrication process, straightforward integration with CMOS technology and broadband (0.2-0.8eV) operation\cite{Sze2006}. The main disadvantage is the limited IPE quantum yield, i.e the number of carriers emitted to Si divided by the number of photons absorbed in the metal, typically$<1\%$\cite{GoykhModel2014, BrongNN2015}. This is mainly due to the momentum mismatch between the electron states in the metal and Si, resulting in specular reflection of "hot" carriers upon transmission at the metal-Si interface\cite{GoykhModel2014, BrongNN2015}. The quantum yield is often called internal quantum efficiency (IQE)\cite{Sze2006}, so that IQE=$I_{ph}/P_{abs}\cdot h\nu/q$, where $P_{abs}$ is the absorbed optical power, $h\nu$ is the photon energy, $q$ is the electron charge and $I_{ph}/P_{abs}$ is the PD responsivity ($R_{ph}$) in units of A/W. One way to improve the IQE, is to confine light at the metal-Si interface by coupling to plasmonic modes\cite{SipeJOSA1981, EndrizAPL1974}. Following this concept, several NIR Si plasmonic Schottky PDs have been demonstrated, exploiting both localized plasmons\cite{FukudaAPL2010, M.KnightSc2011, KwongAPL2012, LinNC2014} and guided surface plasmons polaritons (SPP)\cite{BeriniOE2010, GoykhNL2011, GoykhOE2012, AiharaAPL2011, SobhaniNC2013, DesiatovOpt2015}. Yet, in these device the $R_{ph}$ reported to date does not exceed few tens mA/W with maximum IQE$\sim$1\%\citep{GoykhOE2012}. These values are significantly below SiGe PDs (R$_{ph}\sim$0.4-1A/W and IQE$\sim60-90\%$)\cite{CampbellNP2009, KimerlingNP2010, VlasovNat2010}. Consequently, $R_{ph}$ of Schottky PDs should be further improved both by developing advanced device designs and/or using novel CMOS compatible materials.

Graphene is appealing for photonics and optoelectronics because it offers a wide range of advantages compared to other materials\cite{BonaNP2010, ACF-Roadmap2015, GrigorenkoNP2012, KoppensNL2011, SunACS2010, KoppensNN2014}. A variety of prototype optoelectronic devices exploiting graphene have already been demonstrated, such as transparent electrodes in displays\cite{KimNat2009}, photovoltaic modules\cite{BaugherNN2014,PospischilNN2014}, optical modulators\cite{LiuNat2011, Phare, Ding}, plasmonic devices\cite{ChenNat2012, FeiNat2012, JuNN2011, YanNN2012, EchtNC2012}, and ultra-fast lasers\cite{SunACS2010}. Amongst these, a significant effort has been devoted to PDs, due to a number of distinct characteristics of graphene\cite{BonaNP2010, ACF-Roadmap2015, GrigorenkoNP2012, KoppensNL2011, KoppensNN2014}. Single layer graphene (SLG) is gapless. This enables charge carrier generation by light absorption over a very wide energy spectrum. In addition, SLG has an ultrafast carrier dynamics\cite{BridaNC2013}, wavelength-independent absorption\cite{KuzmenkoPRL2008, NairSc2008}, tuneable optical properties via electrostatic doping\cite{LiNPh2008, WangSc2008}, high mobility\citep{MayorovSc2011}, and the ability to confine electromagnetic energy to unprecedented small volumes\citep{GrigorenkoNP2012, KoppensNL2011}. The high carrier mobility enables ultrafast conversion of photons or plasmons to electrical currents or voltages\cite{XiaNN2009, MuellerNP2010}. By integration with local gates, this process is in-situ tuneable\cite{LeeNN2008,XiaNL2009} and allows for sub-micron detection resolution and pixilation\cite{LemmeNL2011}. SLG absorbs 2.3\% of the incident light\cite{KuzmenkoPRL2008, NairSc2008}, which is remarkably high for an atomically-thin material. This is an appealing property for flexible and transparent opto-electronic devices\cite{BonaNP2010}.

The most common SLG PDs exploit the metal-graphene-metal (MGM) configuration, in which a SLG channel is contacted between source and drain electrodes\citep{LeeNN2008,ParkNL2009,MuellerNP2010,XiaNN2009}. MGM devices are easy to fabricate\cite{MuellerNP2010,XiaNN2009}, they are able to operate over a broadband wavelengths range\cite{MuellerNP2010,XiaNN2009} and have demonstrated ultrahigh ($\sim$230GHz)\citep{UrichNL2011} operation speed. However, for visible and NIR wavelengths free-space illuminated MGM PDs have $R_{ph}\sim$few mA/W\citep{MuellerNP2010,XiaNN2009}. This is primarily because of the finite optical absorbtion\citep{KuzmenkoPRL2008, NairSc2008} and limited photoactive area ($A_{photo}$)\citep{MuellerPRB2009}. In the MGM configuration, the built-in electric field that separates the photoexcited e-h pairs is localized in very narrow ($\sim$100-200nm)\citep{MuellerPRB2009} regions next to the edges of the SLG-metal contacts, whilst the rest of the SLG channel area does not contribute to $I_{ph}$. One way to increase $R_{ph}$ is to apply a voltage between source-drain electrodes and increase the electric field penetration into the SLG channel\citep{XiaNN2009, MuellerNP2010}. However, this will drive a current into SLG (dark current, $I_{dark}$), which could be of the same order or even larger than $I_{ph}$\cite{XiaNN2009, MuellerNP2010}. Thus, this approach can significantly reduce the signal-to-noise ratio (SNR) and increase power consumption. Another way consists in combining MGM devices with metal nanostructures\citep{EchtNC2012, EchtConMat2015} and enabling light coupling to localized and SPP modes, thus enhancing light-graphene interaction and light absorption. MGM-PDs can be also integrated with microcavities\cite{FurchiNL2012, EngelNC2012}, where at resonance the optical absorption in graphene is amplified by multiple light round trips\citep{FurchiNL2012, EngelNC2012}. High $R_{ph}$ can be achieved using a hybrid configuration, in which a MGM structure is combined with semiconductor quantum dots (QD) as light absorbing media\cite{KonstNN2012}. This gave R$_{ph}\sim10^7 A/W$\cite{KonstNN2012} with a photoconducitve gain (i.e. the number of detected charge carriers per single incident photon, $G_{ph}$) up to $10^7$. Similar performances to graphene-QD hybrid devices were also demonstrated in graphene tunneling PDs\cite{LiuNN2014}, comprising two SLGs separated by a thin ($<10$nm) dielectric layer. However, in both QDs-integrated or tunneling-based PDs the typical response time is limited to ms\cite{KonstNN2012,LiuNN2014}, not suitable for high-speed (tens of GHz) optical communications.

Another important performance metric of PDs is the Normalized Photo-to-Dark-current Ratio, NPDR=$R_{ph}/I_{dark}$\cite{ChuiPTL2003}. The larger the NPDR, the better PD noise rejection and ability to perform when a interference (noise) is present. To achieve higher NPDR, $I_{dark}$ must be reduced and $R_{ph}$ must be increased. However, since SLG has no gap, a trade-off between improving $R_{ph}$ by using source-drain bias and minimizing $I_{dark}$ exists in all MGM-PDs\cite{KoppensNN2014}. In telecom applications, where power consumption and SNR are parameters of great importance for achieving energy efficient data transmission with reduced errors rate\cite{ReedSiPhoton2008}, MGM-PDs should be operated near zero bias, which in turn limits $R_{ph}$. Even though MGM-PDs can perform in photovoltaic mode at zero bias with zero dark current\cite{GulbaharIEEE2012, KoppensNN2014}, the conductance of graphene can lead to enhanced thermal noise as a result of reduced channel resistance\cite{GulbaharIEEE2012}. A promising route to increase $R_{ph}$, while minimizing $I_{dark}$, is to create a Schottky junction with rectifying characteristics (i.e a diode) at the SLG-Si interface\cite{ChenNL2011, AnAPL2013, AnNL2013, AmirJQE, WangNP2013}. By operating a Schottky diode in reverse bias (photoconductive mode), $I_{dark}$ is suppressed compared to $I_{ph}$, while the entire Schottky contact area contributes to photodetection\cite{ChenNL2011, AnAPL2013, AnNL2013, AmirJQE, WangNP2013}.

Several PDs have been reported to date, operating at telecom wavelengths and integrating on-chip graphene with Si photonics, based on MGM structures evanescently coupled to Si waveguides\cite{WangNP2013, GanNP2013, PospischilNP2013, YoungbloodNL2013, EnglundNL2015}. In these cases, the guided mode approach enables longer interaction between SLG and the optical waveguide modes than free-space illumination\cite{KoppensNN2014}. This raises the optical absorption in PD beyond $2.3\%$ and, by increasing the interaction length, 100\% light power can be absorbed and contribute to $I_{ph}$\cite{PospischilNP2013}. Nevertheless, because of the evanescent coupling, the typical length needed to achieve nearly complete absorbtion in MGM-PDs is$\sim 40-100\mu m$\cite{WangNP2013,GanNP2013,PospischilNP2013,YoungbloodNL2013,EnglundNL2015}. However, for on-chip optoelectronic integration, where scalability, footprint and cost play an important role, the development of miniaturized, simple to fabricate, Si-based PDs for telecoms, with $R_{ph}$ comparable to the SiGe devices currently employed in Si photonics, is needed\citep{ReedSiPhoton2008}.

Here, we report a compact (5$\mu$m length), waveguide integrated, plasmonic enhanced metal/ graphene/Si (M-SLG-Si) Schottky PD with $R_{ph}\sim$0.37A/W at $1.55\mu m$. The M-SLG-Si structure supports SPP guiding and benefits from optical confinement at the Schottky interface. Our data show that graphene integration in M-SLG-Si PDs increases $R_{ph}$ by one order of magnitude compared to the standard M-Si configuration without SLG. The SLG-integrated device has $R_{ph}\sim85mA/W$ at 1V reverse bias, with $I_{dark}\sim20nA$. Taking advantage of the Shottky diode operation in the reverse bias, $R_{ph}$ can be further increased up to$\sim0.37A/W$ at 3V. To the best of our knowledge, this is the highest $R_{ph}$ reported so far for waveguide-integrated Si-PDs operating at 1.55$\mu$m, and it is comparable to state-of-the-art SiGe devices\cite{CampbellNP2009, KimerlingNP2010, VlasovNat2010}. This is a simple, easy-processed approach for high responsivity Si PDs in the telecom spectral range, and paves the way to graphene-Si optoelectronic integration.

\begin{figure}
\centerline{\includegraphics[width=90mm]{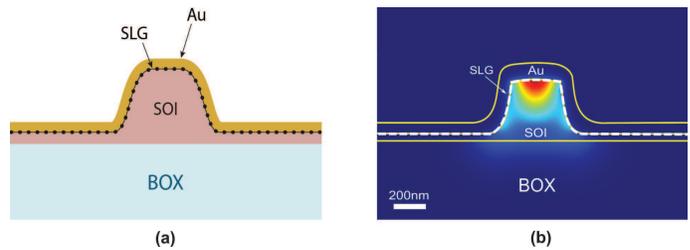}}
\caption{a) Schematic drawing of M-SLG-Si Schottky PD. SOI: silicon-on-insulator. BOX: buried oxide; b) Finite element (COMSOL Multiphysics)\citep{Comsol} simulated optical intensity profile of a SPP waveguide mode supported by a M-SLG-Si structure.}
\label{fig.1}
\end{figure}

Our PD is schematically shown in Fig.\ref{fig.1}a. The device consists of a Si-waveguide coupled to a SLG/Au contact that electrically forms a Schottky diode, while supporting the fundamental SPP waveguide mode (Fig.\ref{fig.1}b). The SPP guiding provides optical confinement at the M-SLG-Si interface (Fig.\ref{fig.1}b), where the IPE process takes place. This can increase light absorption in SLG, adjacent to the Schottky interface and, as a result, enhance $R_{ph}$.
\begin{figure}
\centerline{\includegraphics[width=90mm]{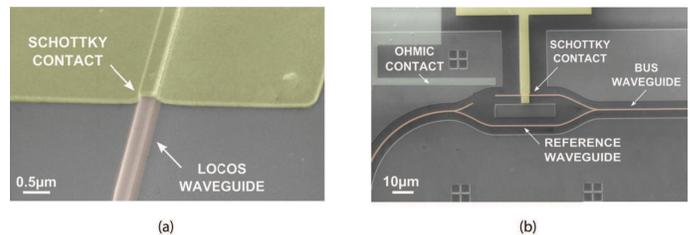}}
\caption{a) SEM micrograph of Schottky PD coupled to Si photonic waveguide. False colors: brown-Si, yellow-Au; b) Layout of waveguide integrated Schottky PD.}
\label{fig.4}
\end{figure}
\begin{figure}
\centerline{\includegraphics[width=70mm]{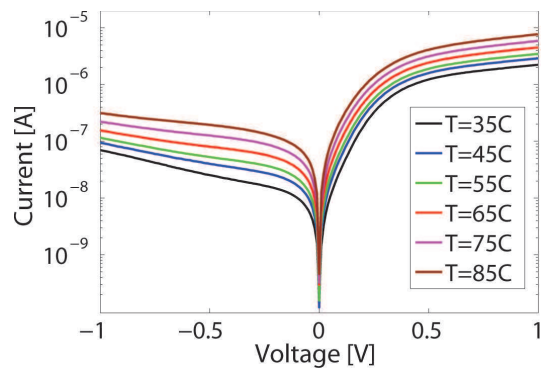}}
\caption{I-V characteristics of our M-SLG-Si Schottky PD for various temperatures.}
\label{fig.5}
\end{figure}

The fabrication process is discussed in Methods. We prepare on the same chip two types of devices: 1) M-SLG-Si Schottky PDs (our target device) and 2) a reference M-Si PD for comparison. Fig.\ref{fig.4} shows a scanning electron microscope (SEM) micrograph of the resulting M-SLG-Si Schottky PD integrated with locally-oxidized\citep{DesiatovOE2010} Si waveguides. The PD length is$\sim5\mu m$ and the Si waveguide width is$\sim310nm$.
\begin{figure*}
\centerline{\includegraphics[width=160mm]{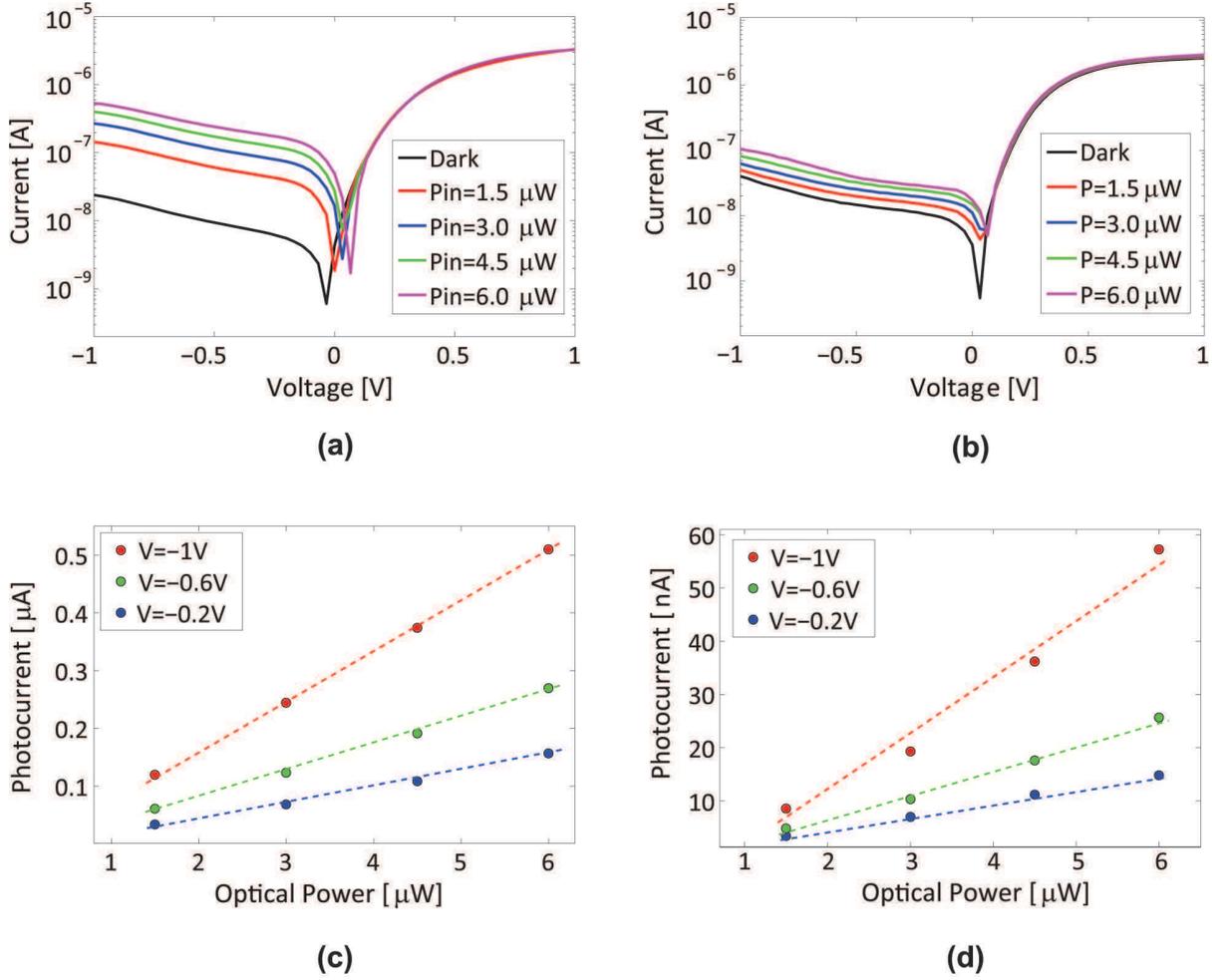}}
\caption{I-V characteristics of a) graphene-integrated and b) reference M-Si PDs for different optical powers coupled to the Schottky region. Measured photocurrent in c) graphene-integrated and d) reference M-Si PDs as a function of optical power coupled to the Schottky region. The slope of the lines in (c,d) corresponds to $R_{ph}$.}
\label{fig.6}
\end{figure*}

Fig.\ref{fig.5} plots a typical current-voltage (I-V) characteristic of our devices, measured using a probe station and a parameter analyzer (Keithley 4200). The device shows electrical rectification (e.g diode behavior). The current in forward bias is limited by series resistance\cite{Sze2006}, while at reverse bias the leakage current $I_0$ is limited by thermionic emission from Au/SLG to Si. In reverse bias, $I_0$ grows with increasing temperature, consistent with what expected for thermionic-emission in a Schottky diode\cite{Sze2006}. In the thermionic regime, the variations of $I_0$ are reflected in the forward bias region, where the forward current also increases (Fig.\ref{fig.5}). Using the I-V characteristics in forward bias, and following the procedure described in Ref.\citenum{CheungAPL1986, SatoJAPL1985}, we extract the M-SLG-Si devices Schottky barrier height $\Phi_B\sim0.34$ and a diode ideality factor n$\sim1.8$ (defined as the deviation of the measured I-V curve from the ideal exponential behavior)\cite{Sze2006}. For the reference M-Si devices we get $\Phi_B\sim0.32$ and $n\sim1.7$, similar to M-SLG-Si. This indicates that SLG does not significantly affect the electrical properties of the Schottky contact.
\begin{figure*}
\centerline{\includegraphics[width=150mm]{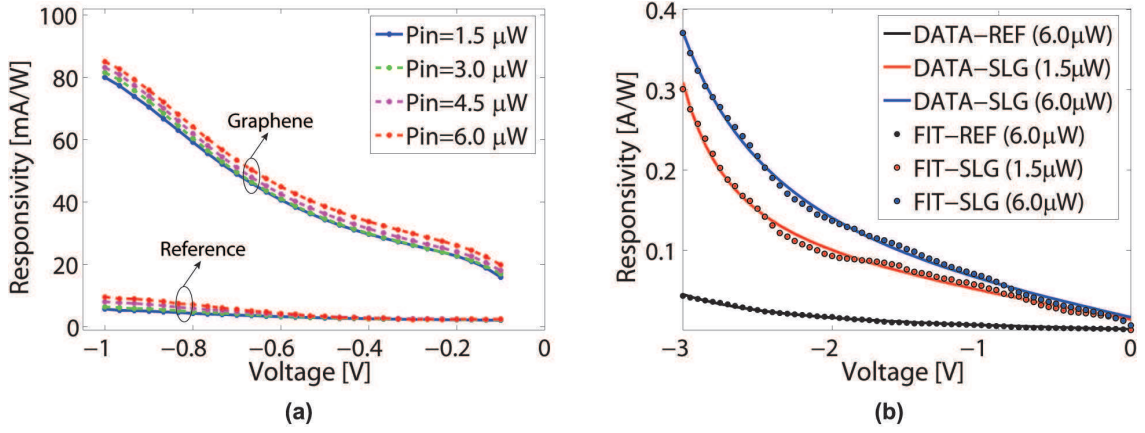}}
\caption{a) $R_{ph}$ of M-SLG-Si and reference M-Si PDs as a function of reverse bias for different optical powers coupled to the Schottky region; b) $R_{ph}$ of M-SLG-Si and reference M-Si PDs for $0<V_R<3V$. Colored solid lines show a fit of the bias dependent $R_{ph}$ based on combined thermionic-field emission and avalanche multiplication processes.}
\label{fig.7}
\end{figure*}

For opto-electronic characterization, we use $1.55\mu m$ transverse-magnetic (TM) polarized light from a tunable laser source (Agilent 81680A) butt-coupled to the waveguide using a polarization-maintaining (PM) tapered fiber. At the output facet of our waveguide the light is collected with a similar fiber and detected by an external InGaAs power meter (Agilent 81634a). Fig.\ref{fig.5} shows that our device has a symmetric Y-branch to split the optical signal between the active arm with integrated Schottky PD and the reference waveguide. This is continuously monitored to avoid optical power fluctuations. To test the opto-electronic response, we measure the I-V curves of graphene-integrated M-SLG-Si and reference M-Si devices at different $P_{opt}$ inside the SPP waveguide, as shown in Fig.\ref{fig.6}a,b. The PDs operate in photoconductive mode\citep{Sze2006}, when a $P_{opt}$ increase results in larger reverse current, since $I_{ph}$ acts as an external current source added to
the Schottky diode $I_0$ .

Fig.\ref{fig.6}c,d plot $I_{ph}$ as a function of $P_{opt}$ as derived from the I-V curves in Fig.\ref{fig.5}. $I_{ph}$ grows linearly with $P_{opt}$ and the slope corresponds to $R_{ph}$, i.e $I_{ph}=R_{ph}\cdot P_{opt}$. We estimate $P_{opt}$ inside the Schottky PDs by taking into account a coupling loss of $\sim18.5dB$ (98.5\%) between the external tapered fiber and the Si waveguide (as measured by monitoring the output signal in the reference waveguide), a propagation loss (scattering + free carriers)$\sim1.5dB/mm$ (29\% per mm) in the waveguide,$\sim3dB$ (50/50) power splitting, and$\sim1.5dB$ (29\%) power loss in the Y-branch. Consequently, based on our I-V measurements and our $P_{opt}$, we calculate and plot $R_{ph}$ as a function of reverse voltage $V_R$ in Fig.\ref{fig.7}a. We get $R_{ph}\sim 85mA/W$ with $I_0\sim20nA$ at $V_R=1V$. The former corresponds to $IQE\sim7\%$. For the reference M-Si PDs we get $R_{ph}\sim 9mA/W$ (at $V_R=1V$), similar to state of the art Si Schottky PDs at $1.55\mu m$\cite{BeriniOE2010, GoykhNL2011, GoykhOE2012, AiharaAPL2011, SobhaniNC2013, DesiatovOpt2015}. We conclude that the presence of SLG at the Schottky interface improves $R_{ph}$ by one order of magnitude compared to our reference M-Si PD. We attribute this to light absorption in the SLG adjacent to the Schottky barrier, where the IPE process takes place. The absorption is enhanced by SPP optical confinement at the M-SLG-Si interface (Fig.\ref{fig.1}b). The significant increase of $R_{ph}$ in SLG-integrated devices could be due to an higher transmission probability of "hot" carriers from SLG to Si when compared to the M-to-Si photoemission process.
\begin{figure*}
\centerline{\includegraphics[width=160mm]{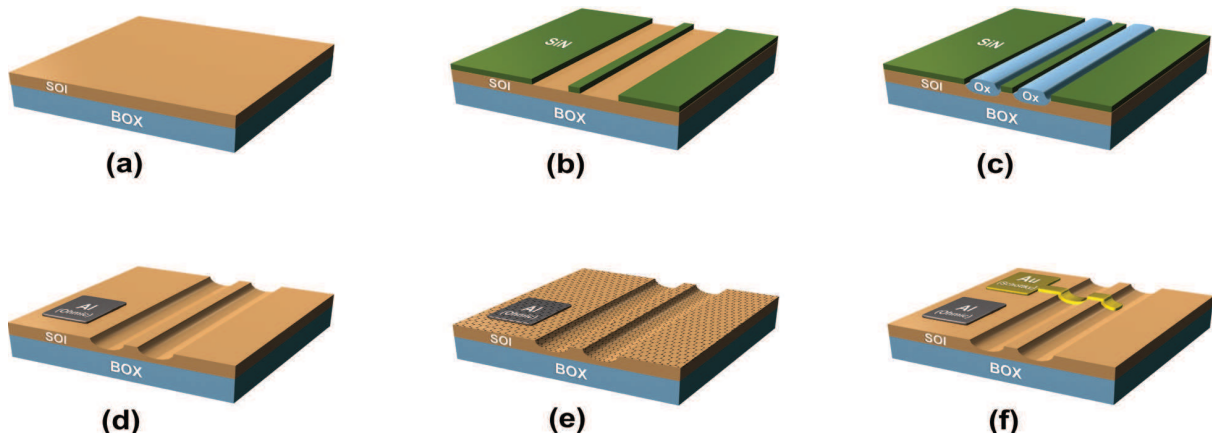}}
\caption{Fabrication process of Si-SLG Schottky PD integrated with a photonic waveguide. a) Planar SOI substrate; b) PECVD deposition and patterning of SiN mask; c) Local oxidation; d) Etching of SiN and SiO$_2$. Al ohmic contact to Si; e) SLG transfer; f) Formation of Schottky contact and consequent SLG etching.}
\label{fig.2}
\end{figure*}

We then measure $R_{ph}$ for $V_R>1V$. Fig.\ref{fig.7}b shows that $R_{ph}$ grows monotonically up to $V_R\sim2V$, then abruptly increases to$\sim0.37A/W$ at $V_R=3V$. To the best of our knowledge, this is the highest $R_{ph}$ reported so far for waveguide-integrated Si-PDs at 1.55$\mu$m, and it is comparable to state-of-the-art Si-Ge devices currently employed in Si photonics\citep{CampbellNP2009, KimerlingNP2010, VlasovNat2010}. We attribute this to the combined effect of two processes that can enhance $I_{ph}$. First: thermionic-field emission (TFE), i.e tunneling of photoexcitepd carriers from the M-SLG contact to Si at energies $E_F<E<\Phi_B$. The relative
contribution of TFE with respect to IPE depends on Si doping, operation temperature and the electric field applied to the Schottky junction\cite{Sze2006, PadovaniAPL1986}. TFE tends to dominate at higher ($>10^{18}cm^{-3}$) doping levels\citep{Sze2006, PadovaniAPL1986} and its voltage dependence is$\propto\sqrt{V_R+\Phi_B/E_0}\cdot exp(qV_R/\epsilon^{\prime})$, where $E_0$ and $\epsilon^{\prime}$ are two analytically defined constants\cite{Sze2006, PadovaniAPL1986}. In our device, with Si doping $\sim7\cdot10^{17}cm^{-3}$ at room temperature, we calculate using Eqs.\ref{eqn:3},\ref{eqn:4} (see Methods) $E_0$ and $\epsilon^{\prime}$ to be$\sim1.04V$ and$\sim2.1eV$ respectively\citep{Sze2006}. Second: avalanche multiplication of photoexcitepd carriers inside the Si depletion region, where the electrons (or holes) can lose their energy upon scattering with the Si lattice creating other charge carriers (i.e impact ionization). This process can be empirically modeled by $M=1/[1-(V_R/V_{BD})^k]$\cite{Sze2006}, where $M$ is the avalanche multiplication factor, $V_{BD}$ is the breakdown voltage at which $M$ goes to infinity, and $k$ is a power coefficient that empirically acquires values between $2<k<6$\citep{Sze2006}. As first order approximation, we assume independent contribution of each process. We show in Fig.\ref{fig.7}b that our data is well fitted by $R_{ph}(V)\propto TFE\cdot M$ with $V_{BD}$ and $k$ as free parameters. From the fit we get $V_{BD}\sim 3.75V$ and $k\sim3.2$, corresponding to $M\sim2$ at $V_R=3V$. We note that, under avalanche conditions, the dark current also increases ($\sim3\mu$A), and operation at elevated V$_R$ ($>$2.5V) reveals a trade-off between improving $R_{ph}$ and higher dark current.

In summary, we demonstrated an on-chip, compact, waveguide-integrated metal-graphene-silicon plasmonic Schottky photodetector operating at $1.55\mu m$. The presence of graphene at the Schottky interface significantly improves the PD responsivity. The device has 85mA/W responsivity at 1V reverse bias, corresponding to 7\% internal quantum efficiency. This is one order of magnitude higher compared to a reference metal-Si PD under the same conditions. We attribute this improvement to the combined effect of light confinement and graphene absorption at the metal-graphene-silicon Schottky interface, as well as enhanced carriers injection from graphene-to-silicon as compared to the metal-silicon interface. Avalanche multiplication for higher ($>$2V) reverse biases allows us to reach a responsivity$\sim0.37A/W$, corresponding to a photogain$\sim2$. Our device paves the way towards graphene integrated silicon photonics.

We acknowledge funding from EU Graphene Flagship (no. 604391), ERC Grant Hetero2D, EPSRC Grants EP/K01711X/1, EP/K017144/1, EP/N010345/1, EP/M507799/1, EP/L016087/1

\section{Methods}
\subsection{Si-SLG Schottky PD Fabrication}
Fig.\ref{fig.2} outlines the fabrication process of our devices. We start with a commercial silicon on insulator (SOI, from SOITEC) substrate with a 340nm p-type ($7\cdot 10^{17}$cm$^{-3}$) Si layer on top of a 2$\mu$m buried oxide (BOX). First, a 100nm SiN mask is deposited by plasma enhanced chemical vapor deposition (PECVD, Oxford
PlasmaLab100) onto the SOI substrate at $300^{\circ}$C (Fig.\ref{fig.2}b). Next, a Si photonic waveguide and the PD area are defined by electron beam lithography (EBL, Raith eLine 150) using positive e-beam resist (ZEP 520A). The EBL pattern is subsequently transferred to SiN by RIE (Oxford Plasmalab 100) with a CHF$_3$/O$_2$ gas mixture. Then, the SOI substrate is locally oxidized (wet, $1000^{\circ}$C), to grow a SiO$_2$ layer only in localized patterns defined by EBL where Si is exposed to O$_2$, while at the same time a SiN mask prevents O$_2$ diffusion into the Si in protected areas (Fig.\ref{fig.2}c). After oxidation, the sacrificial SiN mask layer is etched in hot phosphoric acid (H$_3$PO$_4$, $180^{\circ}$), followed by SiO$_2$ removal in a buffered oxide etch (BOE) solution. The ohmic contact to Si is realized by Al evaporation, followed by metal lift-off and thermal alloying at $460^{\circ}$C in a forming gas (H$_2$/N$_2$, 5\%/95\%) environment. This fabrication process is based on the technique of local-oxidation of Si (LOCOS), in which a Si waveguide is defined by oxide spacers\citep{DesiatovOE2010} rather than reactive ion etching (RIE). The LOCOS process enables the realization of low-loss($\sim$0.3dB/cm)\citep{DesiatovOE2010} Si photonic waveguides coupled to a Schottky PD using the same fabrication step.

\begin{figure}
\centerline{\includegraphics[width=90mm]{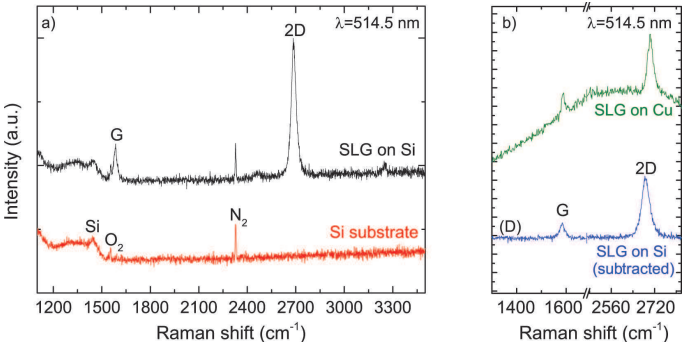}}
\caption{a) Raman spectra of (red curve) Si substrate and (black curve) SLG transferred on Si. b) Raman spectra of (green curve) SLG on Cu , and (blue curve) after normalized, point-to-point subtraction of the Si substrate spectrum (shown in (a) red curve) from the spectrum of SLG transferred on Si (shown in (a) black curve).}
\label{fig.3}
\end{figure}

SLG is grown on a 35$\mu$m Cu foil, following the process described in Ref.\citenum{BaeNN2010}. The substrate is annealed in hydrogen atmosphere (H$_2$, 20sccm) up to $1000^{\circ}C$ for 30 minutes. Then 5sccm CH$_4$ is added to initiate growth\citep{LiSc2009, BaeNN2010}. The substrate is subsequently cooled in vacuum (1mTorr) to room temperature and removed from the chamber. After growth, the quality and uniformity of SLG are monitored by Raman spectroscopy using a Renishaw InVia equipped with a 100X objective (numerical aperture NA=0.85). The Raman spectrum of SLG on Cu at 514nm is shown in Fig.\ref{fig.3}(b) (green curve). This has a negligible D peak, thus indicating negligible defects\citep{FerrPRL2006,CancNL2011,FerrNN2013,FerrPRB2000,BrunACS2014}. The 2D peak is a single sharp Lorentzian with full width at half maximum, FWHM(2D)$\sim$29cm$^{-1}$, a signature of SLG\citep{FerrPRL2006}. Different ($\sim$ 20) point measurements show similar spectra, indicating uniform quality. The position of the G peak, Pos(G), is$\sim$1589cm$^{-1}$, with FWHM(G)$\sim$13cm$^{-1}$. The 2D peak position, Pos(2D) is$\sim$2698cm$^{-1}$, while the 2D to G peak intensity and area ratios, I(2D)/I(G) and A(2D)/A(G), are 2.6 and 5.8, respectively, indicating a p-doping$\sim$300meV\cite{DasNN2008, BaskoPRB2009}, which corresponds to a carrier concentration$\sim5\cdot10^{12}$cm$^{-2}$.

SLG is then transferred onto the SOI with Si waveguides. A$\sim$500nm thick layer of polymethyl methacrylate (PMMA) is spin coated on the SLG/Cu sample, then placed in a solution of ammonium persulfate (APS) in DI water until Cu is completely etched\citep{BaeNN2010, BonaMT2012}. After Cu etching, the PMMA membrane with attached SLG is transferred to DI water for cleaning APS residuals.

To obtain a Schottky interface between the Si waveguide and SLG without the native oxide layer we perform the transfer in diluted hydrofluoric acid (HF) and DI water (1:100) solution. After cleaning from APS residuals, a SLG/PMMA membrane is placed on a plastic beaker containing 5ml/500ml HF and DI water. Next, the target SOI chips are firstly dipped in BOE for 5sec to etch the Si native oxide and then immediately used to lift a floating SLG/PMMA membrane from diluted HF. As a result, during drying the presence of HF at the SLG/Si interface prevents Si oxidation and allows formation of "oxide free" SLG/Si Schottky contacts. After drying, PMMA is removed in acetone leaving SLG to entirely cover the SOI. We also transfer SLG from the same Cu foil using the same transfer procedure onto Si. This is used to check the SLG quality after transfer by Raman spectroscopy.

The Raman spectrum of SLG transferred on Si is shown in Fig.\ref{fig.3}(a)(black line). This is measured at 514.5nm and with laser power below 300$\mu$W to avoid possible heating effects or damage. The D peak region overlaps the bands at$\sim$1200-1500cm$^{-1}$, attributed to third order Raman scattering from TO phonons in the Si substrate\cite{TemplePRB1973}. The peaks at$\sim$1550 and$\sim$2330cm$^{-1}$ in the Raman spectrum of Si substrate (red line) arise from molecular vibrations of ambient oxygen (O$_2$)\cite{WeberJML1960} and nitrogen (N$_2$)\cite{Lofthus1977}. The Raman spectra of the transferred SLG film (black line) and reference Si substrate (red line) are acquired using identical exposure time and laser power. After normalizing the intensity of the third order Si peak at$\sim$1450cm$^{-1}$ in the Si reference spectrum (red line) to the same peak in the spectrum of the transferred SLG film (black line), a point-to-point subtraction is implemented  [Fig.~\ref{fig.3} (b)(blue line)]. The resulting spectrum shows D to G intensity ratio, I(D)/I(G)$\sim$0.04, indicating negligible defects\citep{FerrPRL2006,CancNL2011,FerrNN2013,FerrPRB2000, BrunACS2014}. The 2D peak retains its single-Lorentzian line-shape with FWHM(2D)$\sim$33cm$^{-1}$, validating that SLG has been successfully transferred. Pos(G)$\sim$1584cm$^{-1}$, FWHM(G) $\sim$17cm$^{-1}$ and Pos(2D)$\sim$2687cm$^{-1}$, while I(2D)/I(G) and A(2D)/A(G) are 3.2 and 5.9, respectively, suggesting a p-doping$\sim$4$\cdot$10$^{12}$ cm$^{-2}$ ($\sim$200meV)\cite{DasNN2008, BaskoPRB2009}.

After SLG transfer, we use an additional EBL step followed by O$_2$ plasma etching to selectively remove SLG from the substrate area containing 5 waveguides and dedicated to the reference M-Si devices. Then, a Schottky contact is prepared by evaporation and liftoff of an 3nm/100nm Cr/Au metal strip intersecting the Si waveguide with SLG on top (Fig.\ref{fig.2}f) (or without SLG for reference devices) and forming a Schottky interface for photodetection. Finally, the samples are placed in a reactive $O_2$ plasma, to remove superfluous SLG.
\subsection{Thermionic Field Emission}
The TFE current is given by\cite{Sze2006}:
\begin{eqnarray}
J_{TFE}=\frac{A^{**}T} {k} \sqrt{\pi E_{00}\times q \left[V_R+\frac{\Phi_B}{cosh^2(E_{00}/kT)}\right]} \\ \nonumber
\times exp(\frac{-q\Phi_B} {E_0})\times exp(\frac{qV_R} {\epsilon^{\prime}})\nonumber
\label{eqn:1}
\end{eqnarray}
where $A^{**}$ is the effective Richardson constant, $k$ is the Boltzmann constant, $T$ is the temperature, $q$ is the electron charge. The contribution of TFE to charge injection across the M-Si interface can be evaluated by comparing the thermal energy $kT$ to $E_{00}$, defined as\cite{Sze2006}:
\begin{equation}
E_{00}=\frac {q\hbar} {2} \sqrt{\frac {N} {m^{*}\epsilon_s} }
\label{eqn:2}
\end{equation}
where $\hbar$ is the reduced Planck constant, $N$ is the Si doping, $m^{*}$ is the effective mass of the charge carriers in Si and $\epsilon_s$ is the dielectric permittivity of Si. When $kT\approx E_{00}$ the TFE process mainly contributes to charge carriers injection across the Schottky interface\citep{Sze2006}. The parameters $E_0$ and $\epsilon ^{\prime}$ are analytically defined as\cite{Sze2006}:
\begin{equation}
E_0=E_{00}\times coth(\frac {E_{00}} {kT} )
\label{eqn:3}
\end{equation}
\begin{equation}
\epsilon^{\prime}=\frac {E_{00}} {(E_{00}/kT)-tanh(E_{00}/kT )}
\label{eqn:4}
\end{equation}
In our case, for Si doping $7\cdot 10^{17}$cm$^{-3}$ using Eq.\ref{eqn:2} we get $E_{00}=45meV$, comparable to the thermal energy at room temperature of $26meV$, reflecting a significant TFE contribution to carriers injection at the Schottky interface. Hence, we calculate $E_0$ and $\epsilon^{\prime}$ to be$\sim1.04V$ and $\sim2.1eV$, respectively.

\end{document}